# Three-dimensional composition and electric potential mapping of III-V core-multishell nanowires by correlative STEM and holographic tomography

Daniel Wolf*,†,‡, René Hübner‡, Tore Niermann‖, Sebastian Sturm†, Paola Prete§, Nico Lovergine⊥, Bernd Büchner†, and Axel Lubk†

† Leibniz Institute for Solid State and Materials Research, Institute for Solid State Research, Helmholtzstr. 20, D-01069 Dresden

‡ Helmholtz-Zentrum Dresden-Rossendorf, Institute of Ion Beam Physics and Materials Research, Bautzner Landstr. 400, D-01328 Dresden, Germany

‖ Technische Universität Berlin, Institut für Optik und Atomare Physik, Straße des 17. Juni 135, 10623 Berlin, Germany

§ Istituto per la Microelettronica e Microsistemi, Consiglio Nazionale delle Ricerche, SS Lecce, Via Monteroni, I-73100 Lecce, Italy

⊥ Dipartimento di Ingegneria dell'Innovazione, Università del Salento, Via Monteroni, I-73100, Lecce, Italy

ABSTRACT: The non-destructive characterization of nanoscale devices, such as those based on semiconductor nanowires, in terms of functional properties is crucial for correlating device properties with their morphological/materials features, as well as for precisely tuning and optimizing their growth process. Electron holographic tomography (EHT) has been used in the past to reconstruct the total potential distribution in 3D but hitherto lacked a quantitative approach to separate potential variations due to chemical composition changes (mean inner potential – MIP) and space charges. In this letter, we combine and correlate EHT and high-angle annular dark-field scanning transmission electron microscopy (HAADF-STEM) tomography on an individual <111> oriented GaAs-AlGaAs core-multishell nanowire (NW). We obtain excellent agreement between both methods in terms of the determined Al concentration within the AlGaAs shell, as well as thickness variations of the few nanometer thin GaAs shell acting as quantum well tube. Subtracting the MIP determined from the STEM tomogram, enables us to observe functional potentials at the NW surfaces and at the Au-NW interface, both ascribed to surface/interface pinning of the semiconductor Fermi level.

## Introduction

Nanowires are becoming increasingly important for nanotechnological applications due to their unique electronic, optical, thermal, mechanical and magnetic properties.[1] There is a particularly large variety of nanowire (NW)-based photonic devices including photodetectors, chemical and gas sensors, waveguides, LEDs, microcavity lasers, solar cells and nonlinear optical converters.[2] For example, GaAs–AlGaAs core–shell NWs have great potential for vertically integrated nanolaser sources on a silicon (Si) platform.[3] To understand the NW properties and to monitor and tune the NW synthesis and self-assembly, their structural characterization at the

nanometer and atomic scale using transmission electron microscopy (TEM) techniques is indispensable.[4–6]

Among the various TEM techniques off-axis electron holography (EH) is unique in that it provides access to electro-magneto-static potentials, which are directly related with the electric and magnetic properties.[7–9] Indeed, EH studies on suitable cross-sections of GaAs NWs recently revealed the built-in potential for both axial[10] and radial[11] p-n junctions. While such two-dimensional (2D) imaging techniques can certainly give access to the inner structure of the NWs (e.g. by cross-sectional preparation), they inevitably average in electron beam direction (i.e., projection direction) and suffer from preparational artifacts (e.g., artificial Ga doping by focussed ion beam cutting) in these projections.

To overcome these limitations, electron holographic tomography (EHT), that is, the combination of electron holography and electron tomgraphy, was developed enabling the three-dimensional (3D) reconstruction of the electrostatic potential stripped from projection ambiguities.[12–14] The obtained tomogram, the 3D potential, always contains a contribution from the mean inner potential (MIP), which is defined as the average Coulomb potential within the crystal, where the zero point of the potential is set in vacuum at infinite distance from the crystal.[15] The MIP depends on the material composition, i.e., the atomic species and the crystal structure. Its modulations (e.g. in heterogenous NWs) superimpose those of other potentials to be investigated like the built-in potential or other space charge potentials hence preventing their unique reconstruction and analysis.

In this letter, we overcome this obstacle by independently reconstructing the MIP with the help of high-angle annular dark-field scanning transmission electron microscopy (HAADF-STEM) tomography[16, 17]. We thereby exploit that the HAADF-STEM contrast is sensitive to the atomic number $Z$ hence quantitatively linked to the materials composition in the NW. This allows us to eliminate the MIP contribution from 3D potential reconstructed by EHT and reveal only the space charge potentials, which are related to the NW's electric properties.

We employ this correlated electron tomography analysis to reveal the potential landscape of GaAs–AlGaAs core-multishell NW in 3D. As shown in Figure 1, the latter is fabricated by intergrowing a few nanometer thin GaAs shell in between two thicker AlGaAs shells overgrown around a central GaAs (core) NW. This structure promotes the formation of quantum-confined electron and hole states inside the GaAs shell, thinly wrapped around the hexagonal core [18, 19], resulting in what is called a quantum well tube (QWT). Radial modulation of the NW composition in the form of core-shell or core-multishell NWs adds novel degrees of freedom to the design of NW-based devices with innovative electronic and optical properties. In particular, the combination of photon confinement (waveguiding) and quantum effects, the latter induced by shrinking the shell size below the Bohr exciton radius of the semiconductor gain medium, allows to fabricate novel and tunable nanoscale lasers.[2, 20]

The multi-shell NWs were grown by metal organic vapour phase epitaxy (MOVPE) using Au nanoparticles (NPs) as metal catalyst. In a first step, nearly untapered GaAs NWs, having diameters in the 60–70 nm range, were grown at 400 °C in the form of dense ($10^8$–$10^9$ cm$^{-2}$) arrays on (1 1 1)B-oriented GaAs substrates.[21] These NWs were then radially overgrown at 650 °C by a first Al$_{0.33}$Ga$_{0.67}$As shell, followed by a GaAs shell, a second Al$_{0.33}$Ga$_{0.67}$As shell, and a final GaAs cap layer, the latter to avoid oxidation of the AlGaAs alloy in air. It is worth mentioning here that the central GaAs cores turns out slightly p-type doped,[22] likely as results of unintentional carbon incorporation during the low temperature growth process, while the AlGaAs shells appear n-doped, ascribed to Si contaminations of the Al-alkyl target used for the present growth runs. Detailed analyses of GaAs core excitonic emissions in single GaAs-AlGaAs core-shell NWs further allowed us to estimate the actual net doping concentrations in both the core and AlGaAs shells,[23] which turn out to be around $N_A$=7×10$^{14}$ cm$^{-3}$ and $N_D$=1×10$^{17}$ cm$^{-3}$, respectively. Also, low-temperature photoluminescence measurements have shown GaAs QWT emissions above the GaAs band-gap, and blue-shifting with shrinking of the QWT thickness.[24]

Previously, EHT has been successfully applied to reconstruct electrostatic potentials in GaAs/AlGaAs core-shell NWs, exhibiting sharp interfaces as well as long-range gradients, which are representative of the various self-assembly mechanisms driving the growth of such nanostructures.[25–27] In particular, AlGaAs shells were shown to contain self-assembled Al segregations along <112> directions and local alloy fluctuations arising from the different mobilities of Ga and Al adatoms on the NW sidewall surfaces.

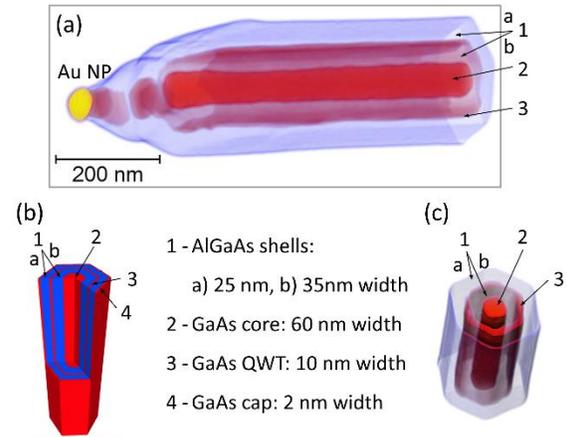

**Figure 1.** 3D structure of GaAs-AlGaAs core-multishell NW grown along <111> orientation by gold catalyst assisted MOVPE. (a) The 3D rendering of the segmented STEM tomogram reveals an interrupted GaAs core in the tapered section. (b) Model of the NW trunk exhibiting its radial distribution. (c) NW trunk cut out from (a) and rotated revealing the experimental radial structure. In the experiments, a continuous 2 nm GaAs cap layer could be observed only partially.

Off-axis EH employs an electrostatic biprism inserted into the electron beam to superimpose the object-modulated image wave with a plane reference wave yielding an interference pattern, the so-called electron hologram, from which amplitude and phase modulations by the object are retrieved quantitatively by Fourier reconstruction method.[28] The phase shift between object wave and unperturbed reference wave caused by a non-magnetic sample can be expressed in the absence of diffraction contrast (dynamical electron scattering) by the phase grating approximation (PGA),

$$\varphi(x, y) = C_E \int_{-\infty}^{+\infty} V_{obj}(x, y, z)\, dz, \quad (1)$$

where $C_E$ is an interaction constant depending on the electron beam energy and $V_{obj}(x, y, z)$ the 3D electrostatic object potential, projected along the beam direction.

Detailed theoretical analysis has proven that in medium (nanometer) resolution the PGA is remarkably well fulfilled.[29] At this resolution, the object potential is mainly given by $V_0$, the MIP of the material. Further contributions are the above-mentioned diffusion potentials caused by free charge carriers, which are also visible at the nanometer scale.

On the other hand, the HAADF-STEM intensity

$$I_{STEM}(x,y) = I_0\left(1 - \exp\left(-\int_{-\infty}^{+\infty} \mu_\alpha(x,y,z)dz\right)\right) \quad (2)$$

can be approximately described as the difference between the initial beam intensity $I_0$ and the bright field intensity, Lambert-Beer-exponentially attenuated with the coefficient $\mu_\alpha$.

The integral in Eq. (2) is often referred to as mass thickness. The attenuation coefficient depends on the type of material and the inner collection angle $\alpha$ corresponding to the inner ring of the HAADF-STEM detector and can be understood as effective density, the product

$$\mu_\alpha(x,y,z) = n(x,y,z)\sigma_\alpha(x,y,z). \quad (3)$$

Here $n(x,y,z)$ is the density of the atoms/scatterers and $\sigma_\alpha(x,y,z)$ the cross-section for electrons scattered into semi-angles higher then $\alpha$, i.e., between the inner and outer ring of the HAADF detector.

Eqs. (1) and (2) describe projections of two 3D physical properties, the object potential and the attenuation coefficient. To reconstruct these physical properties in 3D, a tilt series of projections in different directions, ideally in an angular range of 180°, is recorded in a first step. In a second step, the tilt series is used as input for tomographic reconstruction algorithms, such as weighted back-projection methods[30], simultaneous iterative reconstruction technique (SIRT)[31], or algebraic reconstruction techniques[26]. In HAADF-STEM tomography, the intensities are mostly used directly as projections, which is equivalent to a linear approximation of the exponential function in Eq. (2). However, in our case it turned out that the nonlinear mass thickness dependence in the recorded STEM signal must not be neglected in order to avoid an erroneous amplification of object edges, so-called cupping artefacts, in the STEM tomogram.[32, 33] We eliminated these cupping artefacts by converting the STEM intensity into the mass thickness (Eq. (2)) prior to tomographic reconstruction (see Supporting Information (SI) Sect. I for the details).

The holographic tilt series was acquired at the FEI Titan 80-300 Berlin Holography Special microscope in image-corrected Lorentz mode (conventional objective lens turned off) at an acceleration voltage of 300 kV. Working with 300 kV accelerated beam electrons is preverable over lower acceleration voltages, because scattering absorption, diffraction contrast and inelastic scattering are reduced, hence a better interpretable signal-to-noise ratio is achieved for our 200 nm thick NW specimen. This electron microscope is equipped with two rotatable electron biprisms and an extra lens in between them, permitting an independent adjustment of hologram fringe spacing and field of view without so-called Fresnel fringe artefacts at the hologram borders [34]. We selected a NW sticking with its tip more than one micrometer out over a hole of the lacey carbon support of the TEM grid. The hole was large enough that, by rotating specimen and electron bisprims properly, even at highest tilt angles the carbon support does not disturb object and reference wave.

The HAADF-STEM tilt series was also recorded on an FEI Titan 80-300 microscope operated at an electron acceleration voltage of 300 kV. The camera length used was 160 mm resulting in an angular collection range of the HAADF detector between $\alpha_{min} = 47.5\ mrad$ and $\alpha_{max} = 200\ mrad$. Further parameters for HAADF-STEM and EH tomography tilt series acquisition are summarized in Table 1. Please note in the table that the pixel size of both methods is almost the same, whereas the STEM images have a ca. six times higher resolution than the phase images. The lateral resolution of the phase images is limited by the size of the Fourier mask used to cut out the phase information from the hologram spectrum within the above-mentioned Fourier reconstruction method.

After recording both tilt series, we observed no beam damage or contamination of the NW. We started with the EHT experiment to ensure that the electrostatic potential structure, which we reconstruct by EHT, is related to the pristine NW. During the HAADF-STEM tilt series acquisition, we reduced the total electron dose on the specimen, by blanking the electron beam automatically above the specimen in case that no image is recorded.

The projection data was aligned by coarse correction of displacements between subsequent projections within the tilt series using the cross-correlation method. Fine displacements (sub-pixel) correction and identification of the tilt axis, which is common for all projections of the tilt series, were performed by examining the sinograms involving the center of mass method as described, e.g., in Ref. [8]. Dynamical diffraction may produce global contrast variations in the projections close to zone axis conditions for both the projected potential (EH) and, less strongly, the mass thickness (STEM). These variations are largely removed by normalizing the sum of each projection to the sum averaged over all projections of the tilt series. This operation does not alter the absolute values in the tomogram reconstructed later on from the projections treated that way, but reduces artefacts induced by diffraction contrast (see Supporting Information (SI) Sect. II for the details).

| | STEM tilt series | Holographic tilt series |
|---|---|---|
| **Tilt range** | -68° to +68° | -70° to +71° |
| **Tilt step** | 2° | 3° |
| **Pixel size** | 0.97 nm | 1.09 nm (phase) |
| **Spatial resolution** | 1 nm | 6 nm (phase) |
| **# of projections for tomographic rec.** | 69 | 39 |

Table 1. Parameters for HAADF-STEM and EH tomography tilt series acquisition.

The tomographic reconstruction was computed using weighted SIRT algorithm[35], which involves a weighted back-projection for each iteration step, yielding a faster convergence than a conventional SIRT algorithm. So-called missing wedge artefacts in the tomogram[16] that are caused by the incomplete tilt range available (±70° instead of ±90°) lead to a reduced resolution in the corresponding directions. To minimize this effect, we picked a NW resting with one of its <112> edges (corners of the hexagonal cross section) on the carbon support of the TEM grid. In this particular orientation the faces (edges in the hexagonal cross section) are parallel to the ±60° projection directions and hence well-resolved; in contrast to the orientation where the NW rests flat on a {110} sidewall facet. In addition, we further reduced the missing wedge artefacts at low spatial frequencies by a finite support approach (see Supporting Information (SI) Sect. III for the details).

Having the attenuation coefficient $\mu_\alpha$ and the total electrostatic potential $V_{exp}$ reconstructed, we now proceed with the correlated analysis of both 3D quantities. We first note that the relation between the 3D attenuation coefficient $\mu_\alpha$ reconstructed by HAADF-STEM tomography and the atomic number Z can be approximated by a power law,

$$\mu_\alpha(Z) = a \cdot Z^n, \quad (4)$$

using two free parameters $a$ and $n$. We determined $a = 4.8 \times 10^{-3} \mu m^{-1}$ and $n = 1.8$ by inserting the attenuation coefficients $\mu_\alpha = 2.42 \, \mu m^{-1}$ for GaAs and $\mu_\alpha = 12.32 \, \mu m^{-1}$ for gold, which we obtained from the histogram peaks in the HAADF-STEM tomogram at the NW core and NW catalyst region, respectively (see Supporting Information (SI) Sect. IV for the details). The $Z^{1.8}$ dependence of $\mu_\alpha$, or the related elastic cross-section (Eq. (3)), agrees well with theoretical[36, 37] and experimental values obtained elsewhere (e.g. Krivanek et al.[38]). Furthermore, we can convert the atomic number to the MIP by assuming a linear relation which follows Vegard's law[39, 40] for the intermixing of AlAs and GaAs by

$$V_0(Z) = \frac{V_0^{GaAs} - V_0^{AlAs}}{Z_{GaAs} - Z_{AlAs}}(Z - Z_{AlAs}) + V_0^{AlAs}. \quad (5)$$

Here, the MIP values for GaAs $V_0^{GaAs} = 14.19 \, V$ and AlAs $V_0^{AlAs} = 12.34 \, V$ are taken from density functional theory (DFT) calculations by Kruse et al.[15]. Combining Eqs. (4) and (5) we can finally compute the overall MIP contribution as a function of the attenuation coefficient distribution reconstructed by HAADF-STEM tomography.

In the holographically reconstructed potential, we measure in the GaAs core a value of $V_{exp}^{GaAs} = 13.0 \, V \pm 0.25 \, V$. This value and its deviation are determined by histogram analysis of a $500 \, nm$ long piece at the NW trunc from the corresponding peak position and full width at half maximum (FWHM) (see Supporting Information (SI) Sect. IV for the details). It is smaller than the above-mentioned MIP value of $V_0^{GaAs} = 14.19 \, V$ and experimental values obtained by EH on bulk-like samples[41], but agrees well with previous EHT measurements on GaAs-AlGaAs core-shell NWs[35].

Moreover, the reduction of the measured potential is also consistent with negative charging of the NW under the electron beam illumination, which we can identify by a phase gradient in vacuum indicating electric fringing fields close to the NW (see Supporting Information (SI) Sect. V for the details).

By means of Eqs. (4), (5) and taking into account the negative charging as potential offset of $-1.2 \, V$, we are able to convert the original HAADF-STEM tomogram into a corresponding 3D electric potential distribution. In Figure 2, the different representations of the 3D STEM data, attenuation coefficient (b), atomic number (c) and electric potential (c) are compared with the potential reconstructed by means of EHT (d) on the example of the NW cross-section. The position of the 30 nm thick cross-section is displayed in the volume rendering (a).

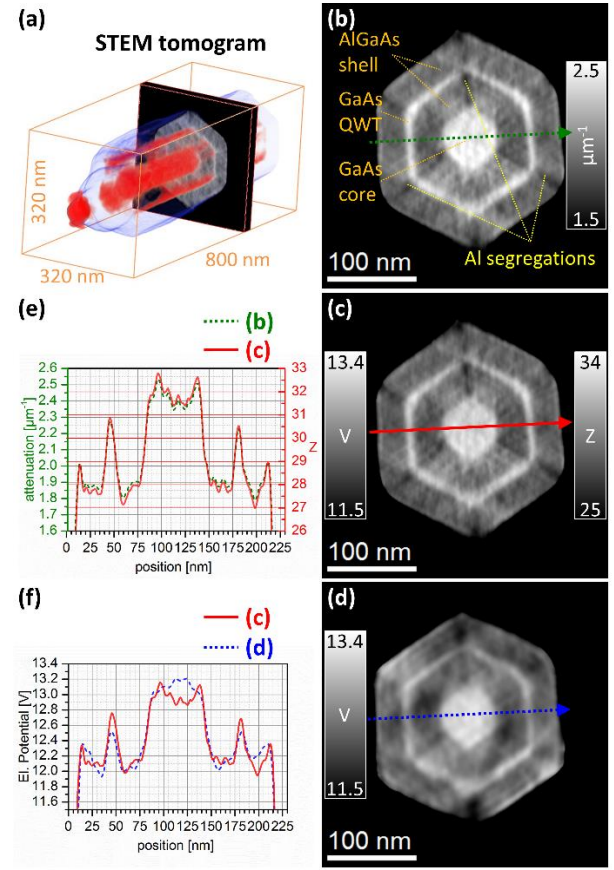

Figure 2. 3D reconstruction of a GaAs-AlGaAs core-multishell NW. (a) Volume rendering of HAADF-STEM tomogram. (b) Cross-section averaged over 30 nm thickness revealing the core-shell structure and the hexagonal shape terminated by the six {110} sidewall facets. The contrast is caused by the different attenuation coefficients $\mu_\alpha$ of GaAs and AlGaAs. (c) Same cross-section as (b), but converted to atomic numbers (Z-contrast) using Eq. (4) and electric potential using Eq. (5). (d) Cross-section through EH tomogram (3D potential) at the same position as (b,c) for comparison. (e,f) Line profiles along the line scans indicated by the arrows in (b), (c), and (d).

Features such as the hexagonal shape terminated by the six {110} sidewall facets, the core-shell structure, and the GaAs QWT are clearly visible in both HAADF-STEM and EH tomogram. However, the Al segregations along the <112> directions, which have also been observed by 2D cross-sectional STEM energy-dispersive X-ray (EDX) measurements [4,42,43], are better resolved in the HAADF-STEM tomogram. We attribute this lower lateral resolution of the EH tomogram to the lower number of projections (phase images) in the tilt series contributing to the tomogram reconstruction (see Fehler! Verweisquelle konnte nicht gefunden werden.). The comparison of the 1D profiles (Figure 2e) reveals that the $Z$-contrast is almost linear with respect to the attenuation signal, which reflects that the above-mentioned $Z^{1.8}$ dependence may be approximated to be linear within the $Z$-range covered by AlGaAs and GaAs. Moreover, the 1D electric potential profiles (Figure 2f) show an excellent match within $0.2\,V$ confirming the quantitative character of both HAADF-STEM and EH tomography. In both cases, the potential in the QWT does not reach the same level as in the GaAs core, suggesting compositional fluctuations within the QWT that coincides with thickness variations.

In order to deduce the aluminum concentration in the NW, we exploit again Vegard's law and adapt it to the relation between the Al concentration $c_{Al}$ and the atomic numbers $Z$ for Al, Ga, and As present in the NW. This allows us to calculate the 3D concentration of Al from the Z-tomogram Z(x,y,z) by

$$c_{Al}(x, y, z) = \frac{2Z(x,y,z) - Z_{As} - Z_{Ga}}{Z_{Al} - Z_{Ga}}. \qquad (6)$$

The result is depicted in Fehler! Verweisquelle konnte nicht gefunden werden.: The longitudinal slice (a) and cross-section (b) present quantitative 2D Al concentration maps displayed in the range from $0.00$ to $0.67$. For example, the longitudinal slice reveals that the QWT is disconnected in the tapered region (see black arrow in Fehler! Verweisquelle konnte nicht gefunden werden.a). In addition, azimuthal variations of the QWT thickness are clearly visible in Fehler! Verweisquelle konnte nicht gefunden werden.b, and measured by line scans 2 and 3. The resulting profiles (Fehler! Verweisquelle konnte nicht gefunden werden.c) provide in case of line scan 2 (solid black line) a thickness (FWHM) of 8 nm and in case of line scan 3 (dashed black line) a thickness of 13 nm. In order to prove that these thickness variations are not caused by missing wedge artefacts, we performed a tomographic reconstruction of a simulated cross-section with constant QWT thickness. To this end, the same tilt range as in the experiment was used. As a result, the tomogram of the simulation does not contain any extra QWT thickness variations induced by missing wedge artefacts (see Supporting Information (SI) Fig. S3 for the details). A more detailed statistical analysis of the GaAs QWT thickness variations reveals that thicknesses between 7 and 12 nm occur in almost equal frequency, but thicknesses down to 4 nm and up to 15 nm are also possible (FWHM values of the thickness variation peak in the histogram, SI Fig. S7). In addition, we observed gaps in the QWT (thickness equal zero), which can be mainly attributed to missing wedge artefacts. (see Supporting Information (SI) Sect. VI for the details). In the cross-section, also the Al segregation is measured (line scan 1) yielding an Al concentration of 65% according to the peak value of the solid red graph in Fehler! Verweisquelle konnte nicht gefunden werden.c. This is in striking agreement with Ref.[4], in which a maximum Al concentration of 62% was measured using spatially resolved EDX spectroscopy. Since we have the entire 3D data available, we can focus our analysis to the overall composition of the AlGaAs shells: the histograms of the AlGaAs shells region for both HAADF-STEM and EH tomogram are shown in Fehler! Verweisquelle konnte nicht gefunden werden.d. They provide a consistent mean Al concentration within the AlGaAs shells in the trunk region of $(41 \pm 12)$ % in case of HAADF-STEM ET, and $(37 \pm 12)$ % in case of EHT. These values are extracted from the histogram peaks (Figure 4d), and their FWHM. Hence, the peak Al concentration are close to the nominal (intended) one of $c_{Al} = 33$% although the measured variation of ±12% is rather large, suggesting a large degree of alloy compositional fluctuation/change within the GaAs shell. We can ascribe ±7% of the variation to a peak broadening that is caused by modulations of the tomogram signal due to tomographic reconstruction artefacts, e.g., mainly missing wedge (see Supporting Information (SI) Sect. VII for the details). Most importantly, these modulations do not shift the peak position as we demonstrate by tomographic reconstruction of a model cross-section while using the same parameters as in the experiment. We further notice that the 65% Al segregation along the NW <112> directions add little contribution to the histograms in Figure 3; on the contrary, they show a tail also towards lower Al contents, ascribable to the growth of Al-poorer AlGaAs regions. Indeed, the formation of Al-poor AlGaAs quantum dots/wires at the apices of {112} facets has been proposed as result of large ratios of mobilities of Al and Ga ad-atoms on {112} facets.[44] However, in our case we do not observe significant Al-poor AlGaAs region at the apices of {112} facets, presumably because these facets are less pronounced at the investigated NW.

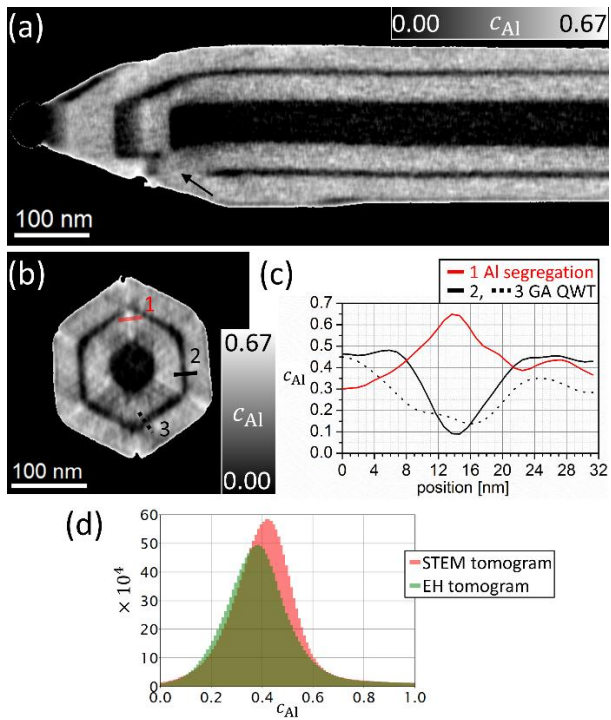

Figure 3. Al concentration mapping within the GaAs-AlGaAs core-multishell NW. Longitudinal slice (a) at position shown in Figure 4a, and cross-section (b) at position shown in Figure 2a are calculated from the corresponding sections of the HAADF-STEM tomogram. (c) Line profiles 1, 2, 3 are taken at the positions indicated in (b). The histograms (d) exhibit Al concentration peaks of 0.41 and 0.37 for the HAADF-STEM and EH tomogram, both measured in the same NW trunk region.

Finally, we exploit the fact that the electric potential which we have computed from the HAADF-STEM tomogram is related to the MIP only, whereas the 3D potential which we have reconstructed using EHT may also contain functional (space charge related) potentials. Consequently, a subtraction of both potentials provides the pure functional potentials. Figure 4 shows longitudinal sections, cut at the same position as Figure 3a, through the 3D potentials obtained from HAADF-STEM (b) and EH tomography (c). They reveal delicate differences: For instance, in the tapered region, we clearly observe a compositional change on the upper edge (green arrows) by accretion of GaAs in the HAADF-STEM obtained MIP, which is not present at the lower edge (blue arrows). In the potential slices obtained from EHT however, increases in the potential of different magnitude, are visible towards both edges. The difference of both potential slices (e), i.e., the subtraction of the MIP contribution yields a symmetric increase in electric potential ($0.5\,V$) on both edges of the tapered region. A similar, albeit weaker, effect can be observed at the NW trunk surface. This is very likely caused by an electric effect associated with the Fermi level pinning at the NW surfaces caused by charged surface states discussed above in the context of fringing fields observed in vacuum. In fact, the negative surface charges, which were also found in a holographic study of Si NWs with a thin surface oxide layer[9], pin the Fermi level in the conduction band. Thus, conduction and valence band are bent downwards corresponding to a potential increase toward the surface in the surface depletion region. 1D simulations based on drift-diffusion model[45] taking into account n-doped $Al_{0.33}Ga_{0.67}As$ shells ($N_D = 1\times10^{17}$ cm$^{-3}$ due to silicon) and slightly p-doped GaAs core and QWT ($N_A = 1\times10^{14}$ cm$^{-3}$ due to carbon) provide the radial distribution of band structure and hence electrostatic potential of the cross-section at the NW trunc (see Supporting Information (SI) Sect. VIII for the details). A comparison of the latter with the experimental cross-section averaged over ca. 200 nm in axial direction revealed agreement within the range of 0.1 V to 0.2 V, which is the accuracy of the experimental data as the histogram data suggest (see Supporting Information (SI) Sect. IV for the details). Furthermore, the line profile (f) in axial direction of the NW reveals potential changes of a few tenths of volts. Most of these changes are difficult to interpret, because differences in sampling and the impact of dynamical scattering in both STEM and EH tomogram may easily lead to variations of such order of magnitude, e.g., at the tip of the GaAs core and its interface to the AlGaAs shell. Notwithstanding, we are convinced that the potential dip of $-0.5\,V$ at the Au-GaAs interface is real (g), because these regions of the tomograms do not suffer from dynamical diffraction, and the difference between the potential profiles is significant (compare red and black curve in (d)). We ascribe this potential slope at the disturbed gold-GaAs interface to Fermi level pinning to "intrinsic" interface states in the middle of the GaAs band gap[46], which is again corroborated by 1D drift-diffusion simulations (blue curve in (g)). Note that the full potential landscape at the tip is a result from drift-diffusion potentials forming between the complex 3D arrangement of the Au particle and the $Al_{0.33}Ga_{0.67}As$ surfaces, which ultimately requires full 3D numerical modelling. In our 1D model, we approximated the impact of the surrounding charged $Al_{0.33}Ga_{0.67}As$ surfaces pinning the Fermi level to the conduction band by pinning the Fermi level to the conduction band within the $Al_{0.33}Ga_{0.67}As$ at a distance of approximately 20 nm to the Au interface.

In total, we have demonstrated how correlated electron tomography techniques, namely electron holographic tomography and high-angle annular darkfield STEM tomography, may be used to reveal the separate distribution of space charge potentials and material-related mean inner potentials. This could only be achieved by a quantitative data treatment and assessment for each processing step throughout the entire workflow of both methods. In our case study of a multiwall GaAs/AlGaAs core-multishell NW, we reveal thickness variations of the quantum well tube in the range of several nanometers as well as an increase of the quantum well width in the tapered region. Both have a profound impact on the nanostructure optical properties, namely an increase of the width of the gain levels in laser applications. Moreover, we observe pinning of the Fermi level at the conduction band for the negatively charged free surfaces of the NW and at

the middle of the GaAs band gap for the interface with the gold catalyst at the tip of the NW

Our correlative characterization of material composition and space charge regions in 3D may be used to adress a variety of questions in modern semiconductor technology, including the design of nanoscale doping profiles, quantum well and quantum dot distributions in novel devices. With our approach it is possible to investigate the relation between morphology and optical as well as electronic properties in a non-destructive manner, which is an important prerequisite for optimizing and advancing their design.

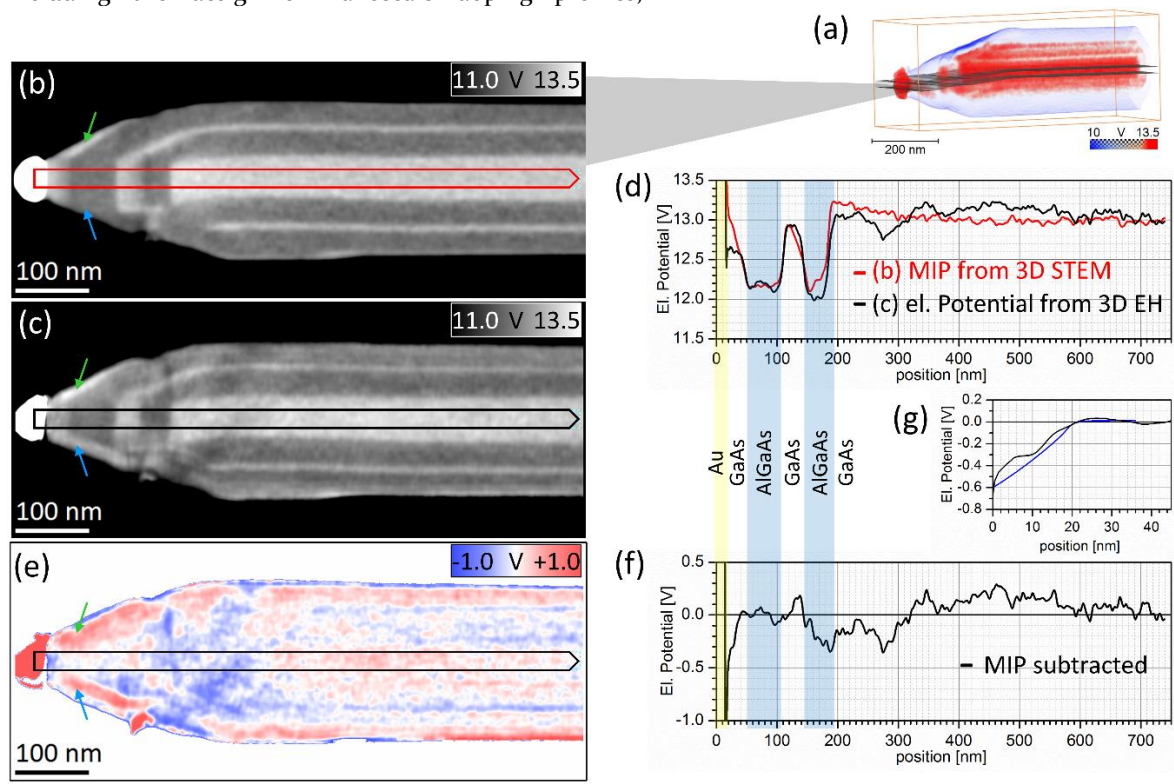

Figure 4. Elimination of the MIP contribution from the electric potential of a GaAs-AlGaAs core-multishell NW reconstructed by EHT. (a) Volume rendering of the MIP distribution gained from HAADF-STEM tomography. (b) Longitudinal slice through the center of the NW at the position indicated in (a). (c) Longitudinal slice through the 3D potential reconstructed by EHT at same position as (b). (d) Profiles at line scans indicated by horizontal arrows in (b) and (c). (e) Electric potential after subtraction of MIP. (f) Profile at the line scan indicated by the horizontal arrow in (e). (g) Zoom in from 0 nm to 40 nm of profile (f) with 1D drift-diffusion simulation indicated as blue line.

## ASSOCIATED CONTENT

**Supporting Information**. The Supporting Information is available free of charge on the ACS Publications website at DOI:

Correction of cupping artefact in the STEM tomogram; Correction of global contrast variations and its influence on tomograms; Finite support approach for missing wedge correction at low spatial frequencies in the tomogram; Histogram analysis of 3D data; Negative charging of GaAs/AlGaAs core-multishell NW during holographic tilt series acquisition; Statistical analysis of GaAs quantum well tube thickness; Influence of missing wedge artefacts on histogram analysis; Comparison of experimental and simulated potentials along cross-sections

## AUTHOR INFORMATION

### Corresponding Author

* Email: d.wolf@ifw-dresden.de

## Notes

The authors declare no competing financial interest.

## ACKNOWLEDGMENT

A.L., S.S. and D.W. acknowledge funding from the European Research Council via the ERC-2016-STG starting grant ATOM. T.N. acknowledges support by Sonderforschungsbereich SFB 787 (subproject A4) of the German Research Foundation (DFG).